\documentclass[aps,prl,preprint,superscriptaddress]{revtex4-2}

\usepackage{graphicx}
\usepackage{dcolumn}
\usepackage{bm}
\usepackage[utf8]{inputenc}
\usepackage[T1]{fontenc}
\usepackage{mathptmx}
\usepackage{etoolbox}
\usepackage{multirow}
\usepackage{blindtext}
\usepackage{array}
\usepackage{amsmath}
\usepackage{microtype}
\usepackage{booktabs}
\usepackage{mhchem}
\usepackage{color}
\usepackage{url}

\begin{document}

\title{Ultrafast Charge Transfer Dynamics at the MoS$_2$/Au Interface Observed via Optical Spectroscopy under Ambient Conditions}

\author{Tao Yang}
\email{tao.yang@uni-due.de}
 \affiliation{Faculty of Physics, University of Duisburg-Essen, 47057 Duisburg, Germany}
 \author{Zhipeng Huang}
 \affiliation{Faculty of Physics, University of Duisburg-Essen, 47057 Duisburg, Germany}
\author{Stephan Sleziona}
 \affiliation{Faculty of Physics, University of Duisburg-Essen, 47057 Duisburg, Germany}
\author{Eckart Hasselbrink}
 \affiliation{Faculty of Chemistry, University of Duisburg-Essen, 45117 Essen, Germany}
\author{Peter Kratzer}
 \affiliation{Faculty of Physics, University of Duisburg-Essen, 47057 Duisburg, Germany}
\author{Marika Schleberger}
 \affiliation{Faculty of Physics, University of Duisburg-Essen, 47057 Duisburg, Germany}
\author{R. Kramer Campen}
 \affiliation{Faculty of Physics, University of Duisburg-Essen, 47057 Duisburg, Germany}
\author{Yujin Tong}
\email{yujin.tong@uni-due.de}
 \affiliation{Faculty of Physics, University of Duisburg-Essen, 47057 Duisburg, Germany}

\date{\today}

\begin{abstract}
To take advantage of the exceptional properties of atomically thin transition metal dichalcogenides (TMDC) for advanced devices and catalysts, integration with metallic surfaces is an efficacious approach for facilitating charge carrier injection and extraction from TMDC monolayers.
Light-matter interactions predominantly occur at the \textit{\textbf{K}} point in TMDC monolayers, making the charge carrier dynamics at this point essential for their optimal performance.
However, direct access to and comprehensive understanding of the charge carrier dynamics at the \textit{\textbf{K}} point of TMDC monolayer on a metal substrate remains challenging.
In this study, we employed azimuth- and polarization-dependent final-state sum frequency generation (FS-SFG) spectroscopy to investigate the ultrafast dynamics of charge transfer at the \textit{\textbf{K}} point of a \ce{MoS2} monolayer interfaced with an Au substrate. 
We observed an ultrafast injection (sub-20 fs) of photoexcited hot electrons from the Au substrate to the conduction band minimum (CBM) of the \ce{MoS2} monolayer.
Subsequently, driven by an internal electric field induced by charge redistribution, injected hot electrons in \ce{MoS2} experience a relaxation and fast return ($\sim2$ ps) from the CBM and a trap state mediated slow return ($\sim60$ ps) process. 
The direct optical observation of the full electron dynamics at the \textit{\textbf{K}} point of \ce{MoS2} monolayer in ambient conditions provides valuable insights into the mechanisms of charge carrier transfer across the TMDC-metal interface, informing the design of advanced TMDC-based devices with enhanced charge transfer rates.
\end{abstract}

\maketitle

\section{Introduction}

Metal-transition metal dichalcogenides (TMDC) monolayer heterojunctions represent a common fabrication configuration aimed at exploiting the distinctive electronic and optical properties of TMDC monolayers for applications in optoelectronics and photocatalysis \cite{wang_electronics_2012,mueller_exciton_2018,sulas-kern_photoinduced_2020}. 
In these applications, metals conventionally act as integral components within electrical circuits and/or function as electron collectors, rendering them indispensable. 
However, compared to the conventional metal-semiconductor junctions found in silicon-based electronics, the metal-TMDC junction confronts several unresolved challenges. 
Foremost among these challenges are high contact resistance (around 1 k$\Omega$ $\mu$m, which is more than 10 times that of silicon-based devices) \cite{chhowalla_two-dimensional_2016,wang_van_2019,shen_ultralow_2021} and an unknown charge transfer rate \cite{xu_weak_2021}, impeding the development of devices characterized by low power consumption and superior performance. 

The high electrical contact resistance predominantly arises from the energy barrier, commonly known as the Schottky barrier, between the metal and the semiconductor. 
This barrier is a consequence of the difference between the work function of the metal and electron affinity of the semiconductor, along with the presence of metal-induced gap states \cite{tung_physics_2014,shen_ultralow_2021}. 
Strategies aimed at mitigating contact resistance involve minimizing the Schottky barrier width by doping \cite{chhowalla_two-dimensional_2016,pelella_electron_2020} or suppressing gap states to attain Ohmic contact \cite{shen_ultralow_2021}. 
The charge transfer rate is pivotal in determining the maximum rate at which a device can inject or extract charge carriers from TMDC monolayers and thus holds particular significance in applications such as field-effect transistors and photodetectors. 

Ultrafast charge dynamics within TMDC monolayers has been extensively studied.
While most studies have concentrated on exciton-related dynamics, encompassing processes such as the formation and recombination of excitons \cite{ceballos_exciton_2016,steinleitner_direct_2017,mueller_exciton_2018,wang_colloquium_2018,madeo_directly_2020,trovatello_ultrafast_2020,dong_direct_2021} and charged-excitons \cite{singh_trion_2016}, there has been relatively less emphasis on the dynamics of free charge carriers. 
Although the optical response in TMDC monolayers is predominantly dominated by excitons, the effective functioning of devices in practical applications requires the dissociation of bound electron-hole pairs into free charge carriers, followed by their transport across an interface, to achieve the desired functionality. 
Given the prevalence of the metal-semiconductor junction in TMDC-based devices, understanding the dynamics of free charge carriers across this interface is critical. 
Earlier investigations have explored the ultrafast dynamics of plasmonic hot carriers across the interfaces between TMDC monolayers and plasmonic metal nanostructures \cite{yu_ultrafast_2016,wen_enhanced_2022,pincelli_observation_2023}. 
However, this configuration may not accurately reflect real-world application conditions when working with standard metal contacts.

As a direct bandgap semiconductor, the bandgap of a TMDC monolayer is typically located at the \textit{\textbf{K}} point, indicating that light-matter interaction predominantly occours at this point in practical monolayer TMDC-based devices.
Consequently, a profound comprehension of the dynamics of free charge carriers injection and/or extraction at the \textit{\textbf{K}} point in TMDC-based devices is imperative as a fundamental prerequisite for realizing broader applications of TMDC monolayers.
Given that the working environment for most TMDC-based devices is normal ambient conditions, studying the charge dynamics under those conditions will significantly improve the knowledge of the real performance of the devices under operation conditions.
Moreover, the potential impact on charge dynamics of a buried fabrication scheme for a TMDC layer is also a relevant consideration since encapsulation of the TMDC layer with a hBN layer is a common strategy to improve the optical quality \cite{wang_colloquium_2018,cadiz_excitonic_2017} and protect the TMDC from surface contamination \cite{cadiz_excitonic_2017}.
A recent study has revealed an ultrafast hot-electron transfer process across the metal-semiconductor interface using time-resolved photoemission electron microscopy (tr-PEEM) \cite{xu_weak_2021}.
However, the probe window of their investigation constrained the dynamics they probed to the vicinity of the $\Gamma$ point in momentum space, which cannot reflect the dynamics at the \textit{\textbf{K}} point due to the distinct band structure. 
Although time- and angle-resolved photoemission spectroscopy (tr-ARPES) enables the investigation of ultrafast free carrier dynamics at the \textit{\textbf{K}} point \cite{cabo_observation_2015}, it is limited to the ultra-high vacuum environment and TMDC surface without additional protective layers.
Consequently, the optical probing of the charge transfer dynamics in a TMDC/metal heterostructure or beyond a buried TMDC interface in ambient conditions remains a significant challenge.

In this study, we employed azimuth- and polarization-dependent final-state sum frequency generation (FS-SFG) spectroscopy to selectively probe the femtosecond resolved charge transfer dynamics at the \textit{\textbf{K}} point within the \ce{MoS2}/Au heterostructure under ambient conditions.
A prior work has demonstrated the ability to isolate the optical response of \ce{MoS2} monolayer from the Au substrate \cite{yang_interaction_2023}.  
A linear optical lineshape showed that the A and B excitons are quenched in a \ce{MoS2} monolayer on Au substrate, and that only free charge carriers are present due to the pronounced dielectric screening and substrate-induced doping effects \cite{yang_isolating_2024}. 
Benefiting from the above findings, the dynamics of free charge carriers can be resolved without the perturbation from excitons. 
By integrating a pump with photon energy below the \ce{MoS2} bandgap, the ultrafast dynamics of hot-electron transfer from the Au substrate across the heterojunction and relaxation at the conduction band minimum (CBM) at the \textit{\textbf{K}} point in the \ce{MoS2} monolayer are unveiled for the first time.

\section{Results and Discussion}
 
\begin{figure}[t]
    \centering
    \includegraphics[scale=0.9]{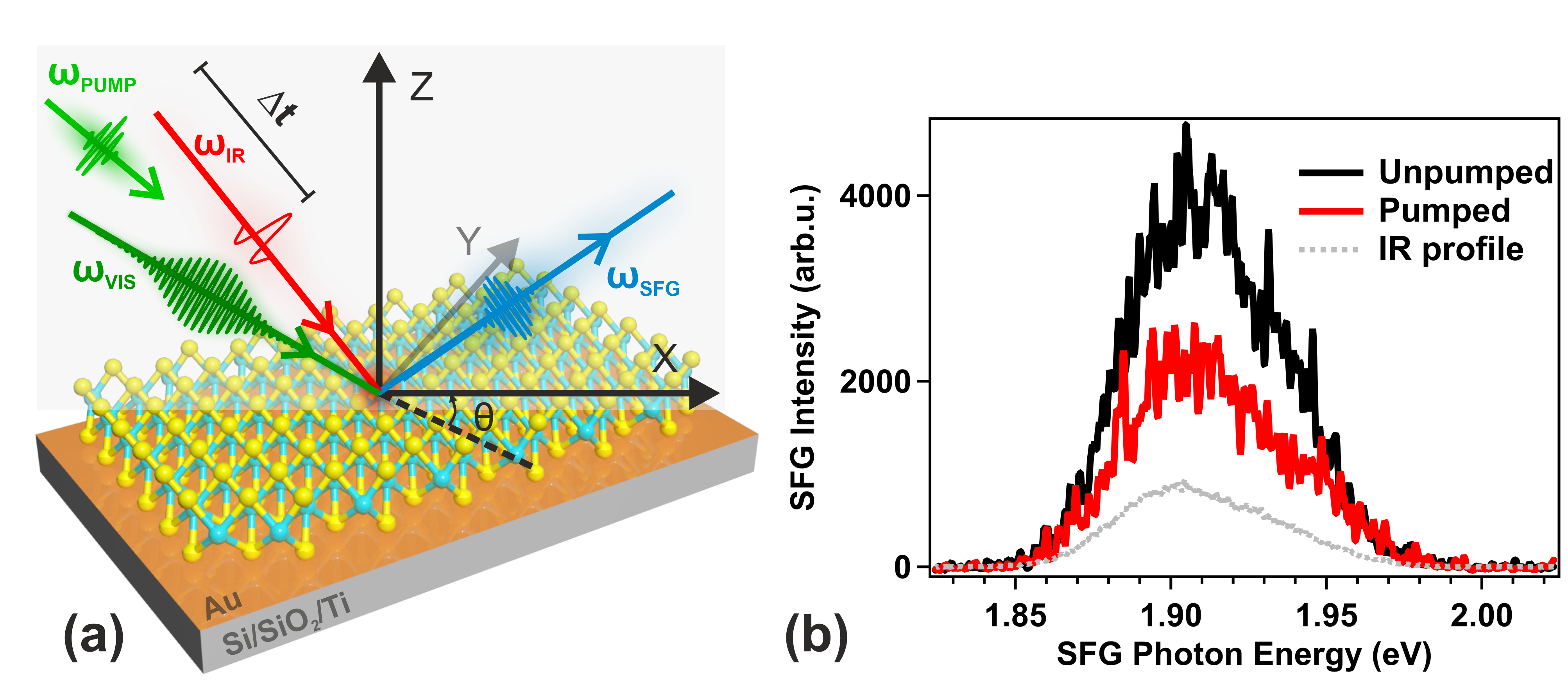}
    \caption{Azimuth-dependent SFG intensity and dynamics for \ce{\ce{MoS2}} monolayer on Au. (a) Schematic representations of an exfoliated \ce{\ce{MoS2}} monolayer on Au substrate investigated by the time- and polarization-resolved FS-SFG. (b) The FS-SFG raw spectra collected before and after the arrival of the pump under \textit{spp} polarization combination. The grey curve represents the raw SFG spectrum on Au substrate under \textit{ppp} polarization combination at equilibrium conditions with offset intensity (divied by 8).}
    \label{fig:setup}
\end{figure}

The preparation of the monolayer \ce{MoS2}/Au heterostructure follows the commonly used mechanical exfoliation method \cite{pollmann_large-area_2021}.
The Au substrate with a thickness of 25 nm was prepared by physical vapor deposition (PVD).
After that, \ce{MoS2} layers were mechanically exfoliated onto the freshly deposited Au substrate to fabricate the \ce{MoS2}/Au heterostructure.
The charge carrier dynamics on \ce{MoS2}/Au were explored by pump-probe FS-SFG spectroscopy using the setup illustrated in Fig. \ref{fig:setup}(a). 
The FS-SFG probe involves overlapping two pulsed incident laser beams, one in the visible and one in the infrared (IR) spectral region, spatially and temporally on the sample surface. 
The resulting emitted sum frequency photons are then detected. 
To realize the time-resolved FS-SFG, an ultrashort pump pulse with different time delays relative to the SFG pulse is introduced to the setup. 
The photon energy of the narrowband visible beam was centered at 1.56 eV, while the photon energy of the broadband IR beam was tuned from 0.28-0.41 eV, ensuring that the final-state resonant SFG photon energies covered the bandgap at the \textit{\textbf{K}} point in momentum space. 
The photon energy of the ultrashort ($\sim$40 fs) pump beam was 1.56 eV, which is smaller than the bandgap of monolayer \ce{MoS2} (1.65 eV) on Au \cite{yang_isolating_2024}. 
The polarization of each beam can be set to either \textit{p} or \textit{s}, allowing for the acquisition of different polarization combinations.
For example, \textit{spp} polarization, where it indicates \textit{s} polarized SFG, \textit{p} polarized visible, and \textit{p} polarized infrared beams.
All experiments were conducted under ambient conditions at $\sim21.5^{\circ}$C. 
For comparison, the dynamics on the Au substrate was also investigated. 
Further details about sample preparation and laser setup can be found in the Supplemental Material.

To investigate the ultrafast charge transfer within the \ce{MoS2}/Au heterostructure, it is necessary to separate the carrier dynamics in the \ce{MoS2} layer from the those within the heterostructure, i.e. to isolate the pure \ce{MoS2} optical response from the heterostructure one.
However, it is always a challenge to isolate the optical response of a TMDC monolayer from the giant background signal of a metal substrate \cite{yang_isolating_2024}.
Second-order nonlinear spectroscopies can overcome this obstacle by taking advantage of their inherent sensitivity to structural symmetry \cite{shen_surface_1989,li_probing_2013}.
As evidenced by various studies on second harmonic generation (SHG) optical response of TMDCs, the 2H-\ce{MoS2} monolayer, which belongs to the $D_{3h}$ point group, has a sixfold azimuth-dependent SHG intensity \cite{kumar_second_2013,li_probing_2013,malard_observation_2013}. 
The different point groups of the \ce{MoS2} monolayer and the Au film result in the emergence of non-zero second-order nonlinear susceptibilities ($\chi^{(2)}$) only when specific polarization combinations of laser beams are employed \cite{yang_interaction_2023}.
By strategically choosing the polarization combination (e.g. \textit{spp} polarization), SFG enables the selective probing of the \ce{MoS2} monolayer and extraction of the pure optical response from the \ce{MoS2} monolayer, free from interference from the Au response.
Furthermore, the isolated pure optical response of the \ce{MoS2} monolayer exhibits a linear lineshape without the commonly observed pronounced A and B exciton features \cite{yang_isolating_2024}. 
A much smaller quasiparticle bandgap was extracted optically, suggesting a significant bandgap renormalization induced by the Au substrate through vigorous dielectric screening and substrate-induced doping. 
Therefore, the quenching of exciton allows us to solely investigate the dynamics of free charge carriers without the influence of excitons.

Figure \ref{fig:setup}(b) presents the FS-SFG spectra of \ce{MoS2}/Au recorded before and after the arrival of the pump, utilizing the \textit{spp} polarization combination. 
The grey curve illustrates the SFG spectrum recorded on the Au substrate with \textit{ppp} (no signal under \textit{spp}) polarization combination under equilibrium conditions. 
The spectra indicate a significant reduction in the FS-SFG signal (bleaching) of \ce{MoS2} following sub-bandgap pumping in comparison to the signal in its equilibrium state. 

\begin{figure}[t]
   \centering
   \includegraphics[scale=0.8]{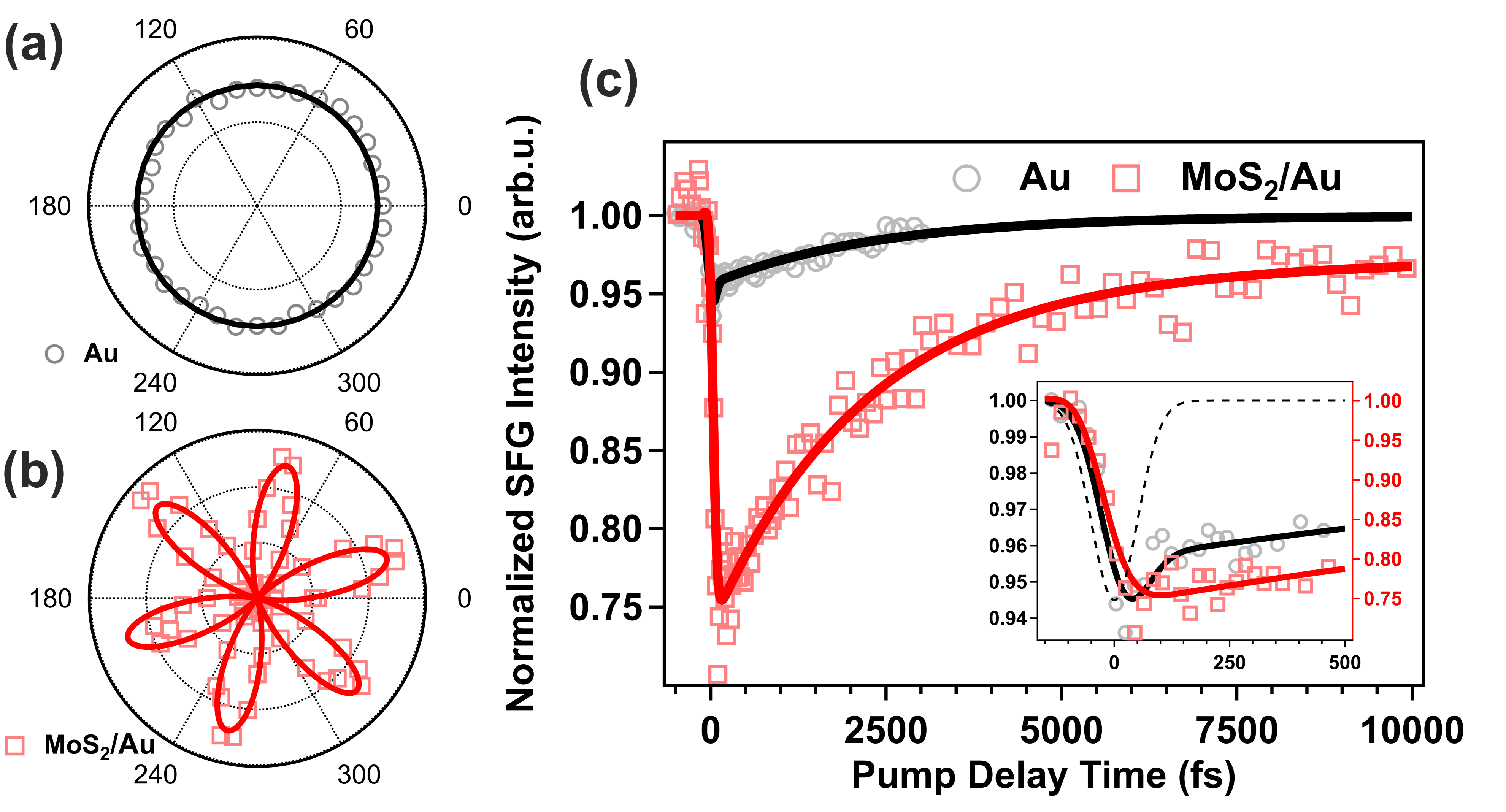}
    \caption{Azimuth-dependent FS-SFG intensity and ultrafast dynamics of the pure Au substrate and the \ce{MoS2} monolayer on Au. (a) Azimuth-dependent FS-SFG intensity of pure Au under \textit{ppp} polarization combination. (b) Azimuth-dependent FS-SFG intensity of \ce{MoS2}/Au under \textit{spp} polarization. (c) Transient signal changes were observed for pure Au and \ce{MoS2}/Au with pumping. The grey circle (black solid line) and red square (red solid line) represent the raw data (fitted curve) of pure Au and \ce{MoS2}/Au under \textit{ppp} and \textit{spp} polarization combination, respectively. The \ce{MoS2}/Au dynamics was collected at one of the azimuthal angles where the FS-SFG intensity is at its maximum. The inset shows the short-term behavior of both samples around SFG signal bleaching maximum with different y-axis scales. In order to make comparison with the Au data, the \ce{MoS2}/Au trace was offset by 60 fs. The black dash line in the inset depicts the simulated instrument response.}
    \label{fig:dynamic}
\end{figure}

Figure \ref{fig:dynamic}(a) and (b) show the azimuth-dependent FS-SFG intensity, which was obtained first to guide spectrum acquisition at different polarization combinations. 
Compared to the isotropic response of pure Au (polycrystalline), the FS-SFG intensity of \ce{MoS2}/Au exhibits a sixfold symmetry, which is consistent with literature reports \cite{kumar_second_2013,li_probing_2013,malard_observation_2013} and our previous studies \cite{yang_interaction_2023,yang_isolating_2024}.
The intensity drops to zero at certain azimuthal angles when using the \textit{spp} polarization combination, suggesting that at this condition the signal is exclusively from the \ce{MoS2} monolayer.
Hence, the pure \ce{MoS2} response was successfully isolated from the \ce{MoS2}/Au heterostructure under \textit{spp} polarization combination. 
For the following experiment, data on the dynamics of \ce{MoS2}/Au were collected at one of the six azimuthal angles where the FS-SFG intensity reaches a maximum. 
In contrast, the dynamics of Au was studied at any azimuthal angle, as there is no azimuthal dependence of Au. 
Since the spectral response of pure Au and \ce{MoS2}/Au are both featureless in our probe energy region \cite{yang_isolating_2024}, to obtain a clearer view of the pump-induced changes in FS-SFG intensity as a function of time delay, we integrated all the SFG spectra at each pump-probe delay time and normalized them to the average spectrum intensity collected at the negative delay time (equilibrium condition), where the probe pulse arrives earlier than the pump pulse. 
The transient signal changes of pure Au and \ce{MoS2}/Au are shown in Fig. \ref{fig:dynamic}(c). 
The y-axis represents the square root of the normalized FS-SFG intensities, which reflects the pump-induced changes in FS-SFG amplitude relative to the one under equilibrium conditions. 

The dynamics of \ce{MoS2}/Au shows a dramatic difference compared to that observed for pure Au at the same pump fluence, as illustrated in Fig. \ref{fig:dynamic}(c). 
The maximum bleaching (largest FS-SFG intensity drop) for pure Au is around 7\%, while a much larger bleaching of approximately 30\% is observed for the FS-SFG signal of \ce{MoS2}/Au.  
Following the maximum bleaching, an ultrafast (within 100 fs) recovery process is observed in the case of pure Au, whereas this ultrafast process is absent in the case of \ce{MoS2}/Au.
After 3 ps, the SFG signal of pure Au recovered to 99\% of its value at equilibrium conditions, while that of \ce{MoS2}/Au is around 92\%. 
Even 100 ps after, the SFG signal of \ce{MoS2}/Au has not fully recovered to its equilibrium conditions (see Fig. S1 for longer dynamics in Supplemental Material). 
A bi-exponential and tri-exponential function convoluted with a Gaussian function is used to fit the data for pure Au and \ce{MoS2}/Au, respectively. 
The $\sigma$ of the Gaussian function was determined to be 50 fs based on the pump-IR cross-correlation trace on Au, a detailed explanation of the procedure we employed can be found in the Supplemental Material. 
Two time constants were extracted for Au with values of $\tau_1$ = 10$\pm$0.1 fs and $\tau_2$ = 2.31$\pm$0.19 ps, corresponding to the well-understood two relaxation processes of hot electrons in metal: electron-electron scattering and electron-phonon scattering \cite{fann_electron_1992,hohlfeld_electron_2000}.

Since the pump photon energy is smaller than the bandgap of \ce{MoS2} monolayer on Au, direct excitation at the band edge of valance band of \ce{MoS2} is impossible. 
However hot electrons can be excited in the Au film of the hetero-contact.
The SFG photon energy we chose here only matches the bandgap of the \ce{MoS2} monolayer located at the \textit{\textbf{K}} point.
Therefore, the observed bleaching and recovery of FS-SFG signal after pumping must be attributed to the dynamics at the \textit{\textbf{K}} point, and the manner in which it is influenced by transferred hot electrons excited from Au. 
By utilizing a Gaussian convoluted tri-exponential fit function, three time constants are extracted for the dynamics of the \ce{MoS2}/Au system: $\tau_1$ = 13$\pm$1 fs, $\tau_2$ = 2.21$\pm$0.17 ps, and $\tau_3$ = 61$\pm$26 ps.
The first time constant (13 fs) describes the buildup of the bleaching of the FS-SFG signal, while the other two (2.21 ps and 61 ps) describe the FS-SFG signal recovery process. 
One interesting observation is that when zooming in on the dynamics around time zero (inset of Fig. \ref{fig:dynamic}(C)), a clearly different trace is observed for pure Au and \ce{MoS2}/Au, i.e., the time to reach the maximum bleaching from the onset of bleaching is delayed for \ce{MoS2}/Au when compared to pure Au. 
In the case of Au, the bleaching of the FS-SFG signal occurred simultaneously with the buildup of the Gaussian function (instrument response), possibly due to the rapidly increased electron temperature after optical excitation.
However, a prolonged bleaching process was observed for \ce{MoS2}/Au. 
As mentioned above, the FS-SFG signal originates from the \textit{\textbf{K}} point of \ce{MoS2} and the direct excitation of electrons in \ce{MoS2} by pumping is impossible.
Therefore, the bleached SFG signal can only be attributed to the occupation of CBM states of the \ce{MoS2} monolayer by hot electrons transferred from Au: the band filling or Pauli blocking effect \cite{liu_engineering_2016,majchrzak_spectroscopic_2021,smejkal_time-dependent_2021}. 
Consequently, the bleaching process (13 fs) of \ce{MoS2}/Au reflects the time interval of hot electrons transfer from Au to the CBM of \ce{MoS2} monolayer.
The fit result suggests a charge transfer time of 13 fs, which is faster than the 120 fs reported in a recent tr-PEEM study \cite{xu_weak_2021}. 
The observed discrepancy in the time constants may be due to a weaker interaction between \ce{MoS2} and the Au substrate. 
Different from the direct exfoliation used in our present work, the tr-PEEM experiment used a stamp to transfer the \ce{MoS2} monolayer, which may result in a larger physical separation between the \ce{MoS2} monolayer and the Au substrate.
The splitting of the $A'_1$ Raman mode can be used as an indicator of the close contact between the \ce{MoS2} monolayer and the Au substrate, as demonstrated in a previous study on our sample \cite{yang_isolating_2024}.

Turning our attention to the FS-SFG signal recovery process, the total recovery time for \ce{MoS2}/Au (longer than 100 ps) is considerably longer than that for Au (within 10 ps). 
As previously stated, the bleaching of the FS-SFG signal is a consequence of the band-filling or Pauli blocking effect. 
Therefore, the recovery of the FS-SFG signal indicates that hot electrons are leaving the CBM and the occupied states are being released. 
The question then arises: where do these hot electrons go?
Given that no holes are generated in \ce{MoS2} during the sub-bandgap pumping, recombination with holes in \ce{MoS2} can be ruled out. 
A portion of excited hot electrons in Au are transferred to \ce{MoS2}, establishing an opposite charge condition on both sides. 
This charge imbalance will give rise to an electric field at the interface, which will act as the driving force for the transfer of hot electrons from the CBM of \ce{MoS2} back to Au, thus achieving charge neutrality. 
The second time constant, about 2 ps, is nearly identical to the time scale of 2.31 ps, during which the hot electrons in Au achieve equilibrium. 
Therefore, the second time scale (2.21 ps) reflects the equilibration of the CBM occupation of \ce{MoS2} with the overall Fermi level of the combined system, \ce{MoS2}/Au, and the process by which hot electrons present in \ce{MoS2} are transferred back to Au.
It is worth noting that the tr-PEEM research proposed a time constant of $\sim400$ fs for the hot electrons back transfer to Au from monolayer \ce{MoS2} conduction band \cite{xu_weak_2021}. 
Due to the conservation of lateral momentum, their study's probe window (range of momenta probed) was constrained to the vicinity of the $\Gamma$ point of \ce{MoS2}. 
Consequently, the 400 fs time constant should be interpreted as the time that hot electrons left their probe window in proximity to the $\Gamma$ point, rather than as a measure of the transfer of hot electrons from the CBM at the \textit{\textbf{K}} point of the \ce{MoS2} back to the Au substrate.
In contrast to this tr-PEEM study, our experiment allows for the direct observation of ultrafast hot-electron dynamics at the CBM of the \textit{\textbf{K}} point. 
This is made possible by the use of the FS-SFG, which probes the direct interband transition at the \textit{\textbf{K}} point of \ce{MoS2} monolayer.

We note that there can be a difference in charge transfer rates between the forward transfer (13 fs from Au to \ce{MoS2}) and the back transfer.
This is due to the fact that the forward transfer is mostly achieved by very hot electrons immediately after excitation in Au, whereas the back transfer occurs via tunnelling of electrons close to the Fermi level.
The Schottky barrier height for \ce{MoS2} monolayer on Au has been determined in various studies to be in the range of 0.5-1.0 eV \cite{kang_computational_2014,lee_layer-dependent_2019,pollmann_large-area_2021}. 
Given that a pump beam with a photon energy of 1.56 eV was employed, the photoexcited hot electrons possess sufficient energy to surpass the barrier when transferring from Au to the \ce{MoS2} monolayer.
However, for the transferred hot electrons to return back to Au, tunnelling through the barrier will be the dominant mechanism, resulting in a somewhat slower back transfer rate.

To corroborate the aforementioned interpretation, the charge transfer between Au and \ce{MoS2} is analyzed using a simple kinetic model. 
This model allows us to elucidate the manner in which equilibrium is established in the combined electronic system of Au and \ce{MoS2}, thereby rationalizing the nearly identical time constants observed in the picosecond range in the pristine Au film and in \ce{MoS2}/Au.  
In brief, the proposed kinetic model posits that the transient FS-SFG response in pure Au is proportional to the electron temperature following pump excitation. 
The maximum temperature of the electrons is calculated based on the two-temperature model \cite{anisimov_electron_1974, hohlfeld_electron_2000}.
In the case of the dynamics observed in the \ce{MoS2} monolayer, the occupation at the CBM in \ce{MoS2} determines the transient FS-SFG response observed in the \ce{MoS2}/Au system.
To describe the occupation of the CBM in \ce{MoS2}, it is necessary to consider the tunneling of electrons from Au to \ce{MoS2} and the back-tunneling process.
The non-equilibrium population of the \ce{MoS2} orbitals near the CBM follows a first-order differential equation, which is solved to give the occupation of the CBM of the \ce{MoS2} monolayer. 
The unoccupied population (1 - population) at the CBM of \ce{MoS2} is illustrated in Fig. S4 in the Supplemental Material.
A fit of the trace suggests a decay time constant of 2.31 ps, which is identical to the $\tau_2$ = 2.31 ps observed in the pure Au.
More details can be found in the section on the kinetic model in the Supplemental Material.

With regard to the slowest recovery process observed for \ce{MoS2} on Au, further evidence from the change in spectral centroid as a function of delay time will facilitate a deeper understanding of this phenomenon.
The spectral centroid, defined as the center-of-mass of a spectrum, was extracted for each spectrum and plotted as a function of pump-probe delay time. The spectral centroid can serve as a reflection of the real-time occupation of states at the CBM of the \ce{MoS2} monolayer.
As states in the CBM of \ce{MoS2} monolayer are occupied, only transitions to higher energy states are allowed, resulting in an increased probability of high-energy transitions and a reduced probability of low-energy transitions.
The direct impact on the spectrum is a blueshift of the spectral centroid.
As illustrated in Fig. S3 of the Supplemental Material, the spectral centroid for \ce{MoS2}/Au initially exhibits a blueshift with respect to that at equilibrium conditions, followed by a return to its equilibrium position.
The observed blueshift in the spectral centroid over time is due to the gradual occupation of CBM states by hot electrons injected from the Au substrate.
The tunneling-out of hot electrons and the subsequent release of occupied states at CBM facilitates the low energy photoexcitation, which is observed as decay of blueshift.
A single exponential fit suggests a time constant of $2.37\pm0.76$ ps for the decay, which is in close agreement with the time required for hot electrons to leave the CBM of \ce{MoS2}. 
This implies, in principle, the complete release of the CBM states and full recovery of FS-SFG signal of \ce{MoS2} monolayer after $\sim$2 ps.
However, in reality, only partial recovery of the FS-SFG signal is observed.
This indicates that, although the injected hot electrons have left the CBM and released the previously occupied CBM states at the \textit{\textbf{K}} point, the screening effect still exists. 
One possible source of the observed additional screening effect ($\sim$60 ps) may be a defect-mediated trap state situated close to the CBM \cite{wang_ultrafast_2015,komsa_native_2015,schuler_large_2019,krause_microscopic_2021,handa_spontaneous_2024}.
The defect can trap a portion of the hot electrons transferred from Au, which results in the long-lasting weak bleaching of the FS-SFG signal due to the Coulomb repulsion between trapped hot electrons and excited electrons from \ce{MoS2}.
Ultimately, the trapped hot electrons will gradually return back to the Au substrate, and the FS-SFG signal will be fully recovered.
Such a trap state, e.g. due to a S-vacancy in the \ce{MoS2} monolayer, was observed by scanning tunneling spectroscopy \cite{schuler_large_2019} and tr-APRES \cite{krause_microscopic_2021} within the bandgap.

\begin{figure}[t]
    \centering
    \includegraphics[scale=0.8]{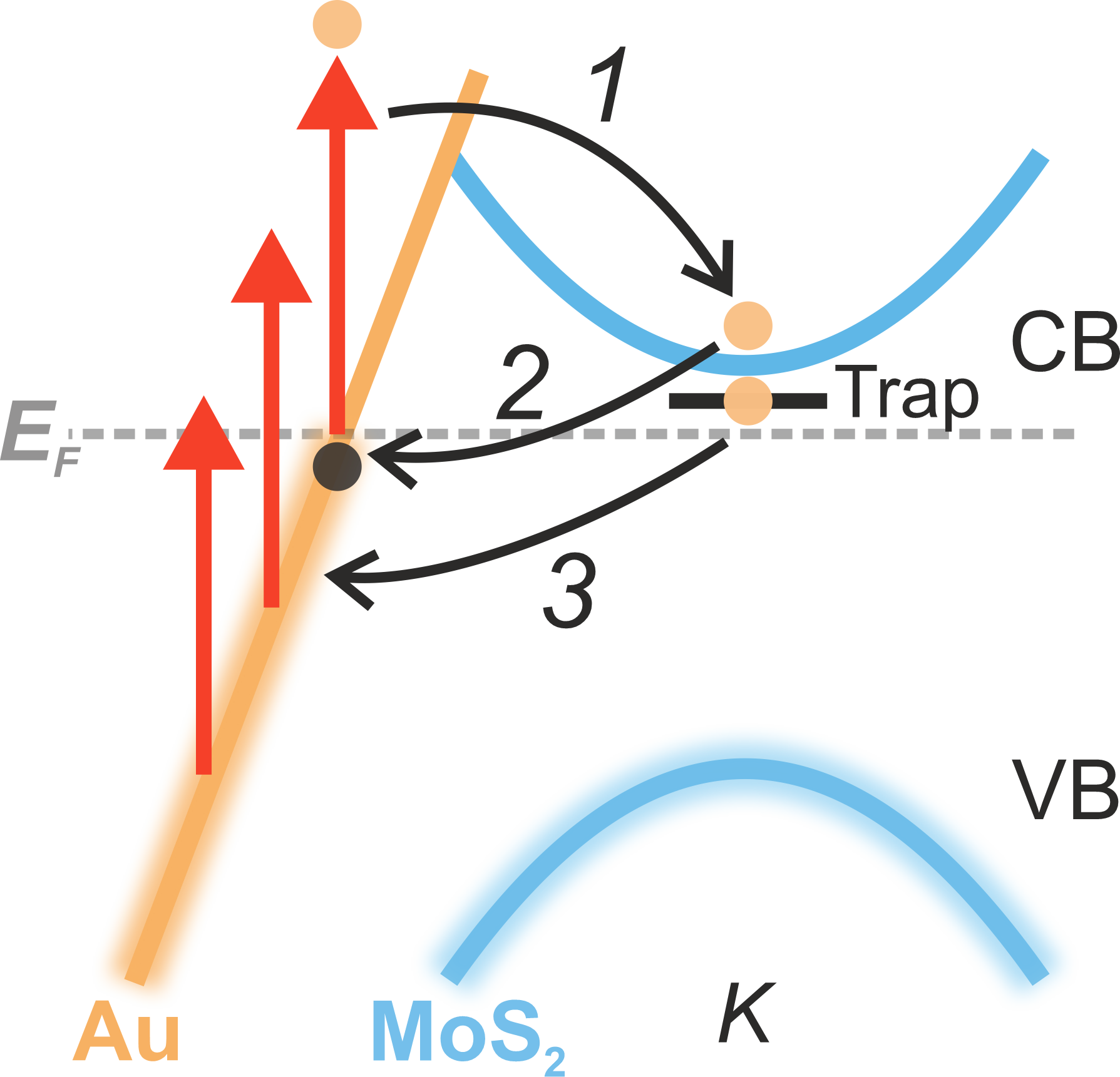}
    \caption{Charge transfer process within the \ce{MoS2}/Au heterostructure, illustrated in a simplified band structure picture of Au and \ce{MoS2} monolayer around the \textit{\textbf{K}} point in momentum space.}
    \label{fig:mechanism}
\end{figure}

Based on the above analysis, we summarize our understanding of the observed dynamics in Fig. \ref{fig:mechanism}. 
At equilibrium, the FS-SFG signal originates primarily from the \ce{MoS2} monolayer ensured by choosing the appropriate polarization combination. 
Upon the arrival of the pump pulse, hot electrons are excited instantaneously in the Au substrate, while the \ce{MoS2} remains in its equilibrium state since the pump photon energy is below the bandgap. 
After less than 20 fs, part of the hot electrons photoexcited in Au will transfer to \ce{MoS2} and reach the CBM, thereby also filling the trap state at the \textit{\textbf{K}} point.
Subsequently, a charge redistribution-induced internal electric field drives the gradual return of hot electrons from the CBM of \ce{MoS2} to Au.
Thereafter, the small amount of hot electrons trapped in the defect state will also return back to Au at a much slower rate, contributing to the second recovery process ($\sim60$ ps).
The present study provides a direct observation of the injection and extraction of hot electrons into the CBM at the \textit{\textbf{K}} point of the \ce{MoS2} monolayer. 
The extracted time constants provide compelling evidence of the underlying mechanisms governing the charge transfer and relaxation processes. 
This not only advances our understanding of charge carrier transfer across the \ce{MoS2}-Au interface but also opens a new avenue for the TMDC community. 
It offers a robust methodology to characterize the ultrafast transfer of charge carriers across heterointerfaces, without being limited to TMDC-metal interfaces, but rather allows for an extension to molecules adsorbed on TMDC or TMDC heterostructures under ambient conditions. 
Furthermore, the implementation of selective circular polarization of pump and probe beams offers additional accessibility to valley-associated dynamics.

\section{Conclusion}

In summary, we have employed polarization- and azimuth-dependent FS-SFG to optically investigate the hot electron dynamics across the \ce{MoS2}/Au interface and within the \textit{\textbf{K}} point of the \ce{MoS2} monolayer under ambient conditions, free from the interference by of excitons.
Compared to the bare Au, significant differences were observed in the dynamics for the \ce{MoS2} monolayer on Au, including a much larger bleaching size and a distinct relaxation process.
A much faster electron transfer rate (sub-20 fs) compared to a previous report was observed at the \ce{MoS2}/Au interface owing to the optimal contact and strong interaction between the two components achieved in this study.
As observed in both the pure Au and the \ce{MoS2}/Au system, the nearly identical relaxation process occurs in $\sim2$ ps, indicating that the equilibrium is established in the combined electronic system, as successfully elucidated by the simple kinetic model.
A defect-mediated process was proposed to account for the slower dynamics of hot electron relaxation extending to around 60 ps.
These findings will facilitate the design of advanced TMDC-based devices with fast charge transfer rates through optimized contacts.

\begin{acknowledgments}
This work was funded by the Deutsche Forschungsgemeinschaft (DFG, German Research Foundation) through projects A06, B02, and C05 within the SFB 1242 "Non-Equilibrium Dynamics of Condensed Matter in the Time Domain" (project number 278162697), through Germany's Excellence Strategy (EXC 2033 - 390677874 - RESOLV). Additional support was provided by the European Research Council, i.e., ERC- CoG-2017 SOLWET (project number 772286) to RKC. 

\end{acknowledgments}


%

\end{document}



\title{Ultrafast Charge Transfer Dynamics at the MoS$_2$/Au Interface Observed via Optical Spectroscopy under Ambient Conditions}
\textbf{\textit{Supplemental Material for}}

\author{Tao Yang}
\email{tao.yang@uni-due.de}
 \affiliation{Faculty of Physics, University of Duisburg-Essen, 47057 Duisburg, Germany}
 \author{Zhipeng Huang}
 \affiliation{Faculty of Physics, University of Duisburg-Essen, 47057 Duisburg, Germany}
\author{Stephan Sleziona}
 \affiliation{Faculty of Physics, University of Duisburg-Essen, 47057 Duisburg, Germany}
\author{Eckart Hasselbrink}
 \affiliation{Faculty of Chemistry, University of Duisburg-Essen, 45117 Essen, Germany}
\author{Peter Kratzer}
 \affiliation{Faculty of Physics, University of Duisburg-Essen, 47057 Duisburg, Germany}
\author{Marika Schleberger}
 \affiliation{Faculty of Physics, University of Duisburg-Essen, 47057 Duisburg, Germany}
\author{R. Kramer Campen}
 \affiliation{Faculty of Physics, University of Duisburg-Essen, 47057 Duisburg, Germany}
\author{Yujin Tong}
\email{yujin.tong@uni-due.de}
 \affiliation{Faculty of Physics, University of Duisburg-Essen, 47057 Duisburg, Germany}

\date{\today}

\maketitle

\section*{\hfil Contents \hfil}

\ref{fig}. Figures 

\ref{details}. Sample preparation, laser setup, and azimuthal-dependent FS-SFG pump-probe measurement

\ref{resolution}. Determine the $\sigma$ of the Gaussian function for fitting

\ref{model}. A simple kinetic model

Reference


\pagebreak

\section{\label{fig}Figures}

\begin{figure}[h]
\renewcommand{\thefigure}{S\arabic{figure}}
    \centering
    \includegraphics[scale=0.5]{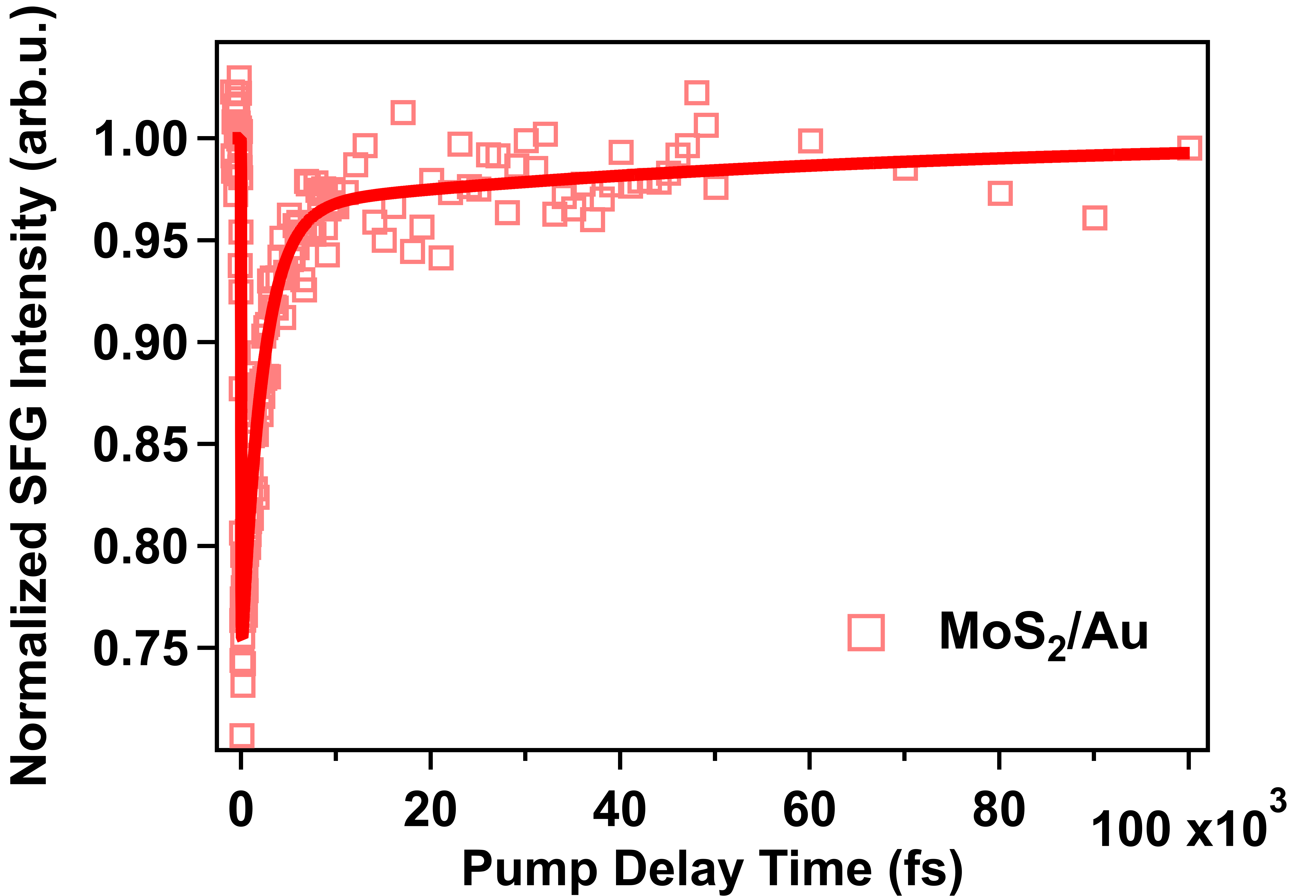}
    \caption{Ultrafast dynamics of \ce{MoS2}/Au under \textit{spp} polarization combination at long timescale. }
    \label{fig:S1}
\end{figure}

\begin{figure}[h]
\renewcommand{\thefigure}{S\arabic{figure}}
    \centering
    \includegraphics[scale=0.5]{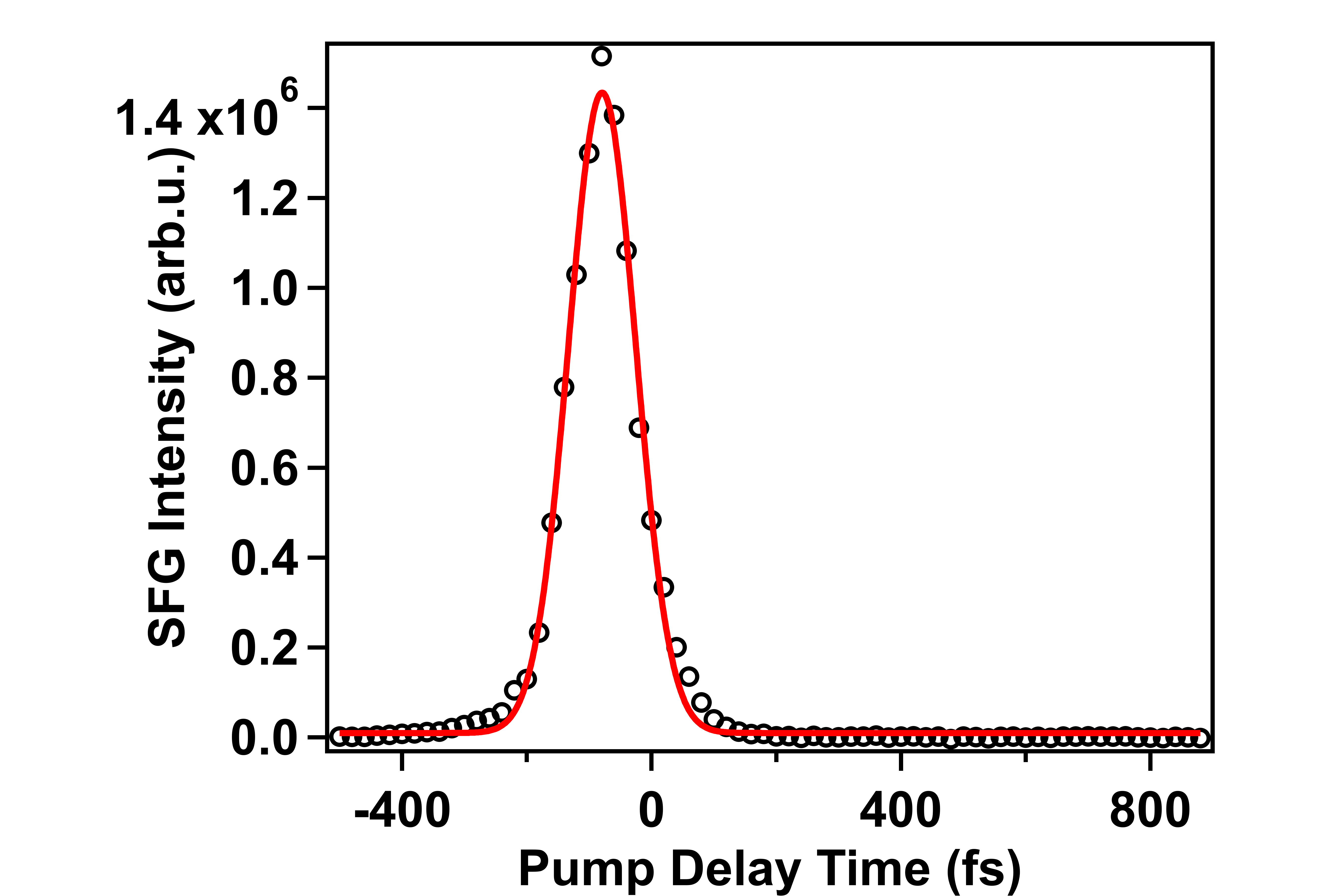}
    \caption{The pump-IR cross-correlation trace measured on Au. The fit suggests a full width at half maximum of 127 fs.}
    \label{fig:S2}
\end{figure}

\begin{figure}[h]
\renewcommand{\thefigure}{S\arabic{figure}}
    \centering
    \includegraphics[scale=0.75]{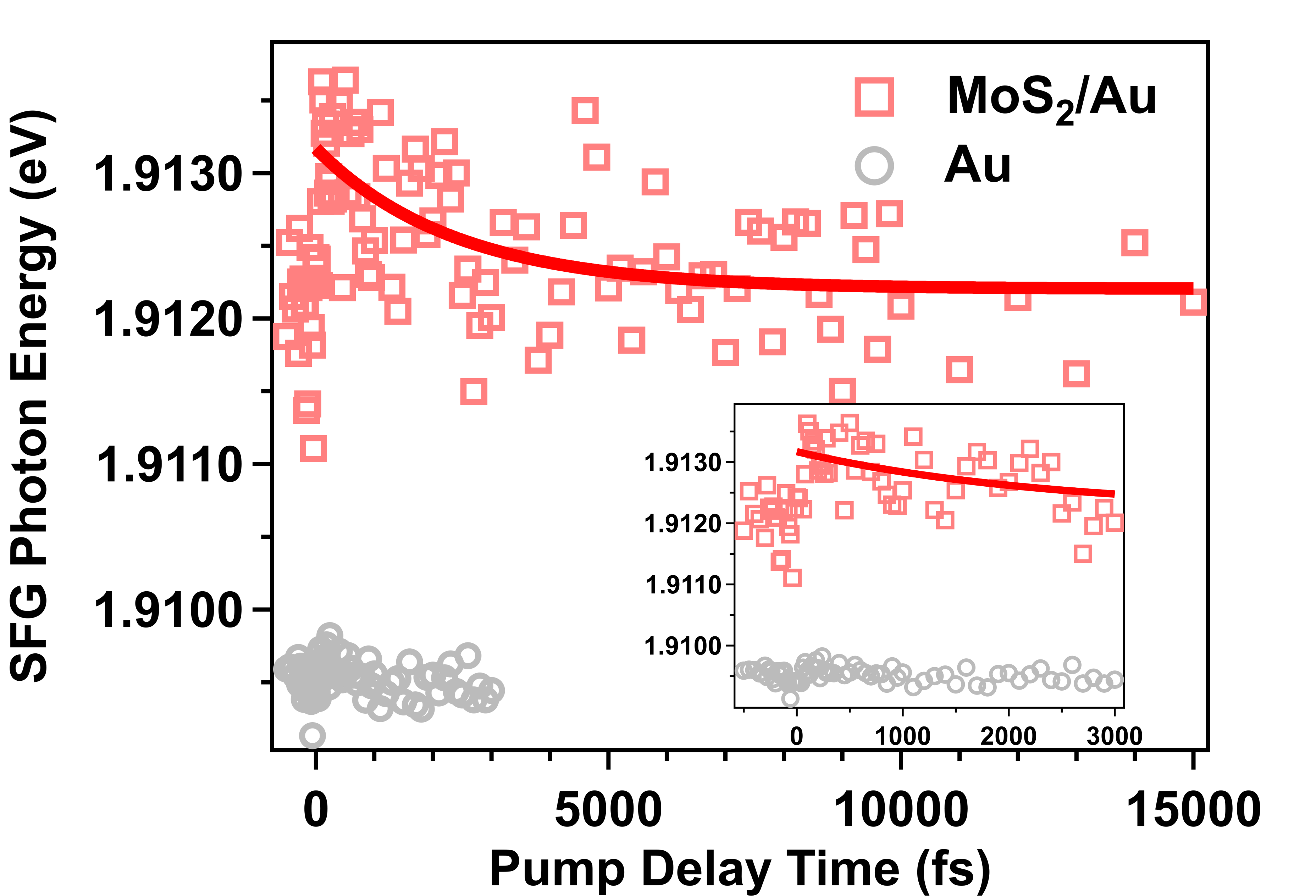}
    \caption{The FS-SFG spectral centroid as function of pump delay time for \ce{MoS2} monolayer on Au and pure Au. Inert shows the zoom in dynamics near time zero.}
    \label{fig:S3}
\end{figure}

\begin{figure}[h]
\renewcommand{\thefigure}{S\arabic{figure}}
    \centering
    \includegraphics[scale=0.5]{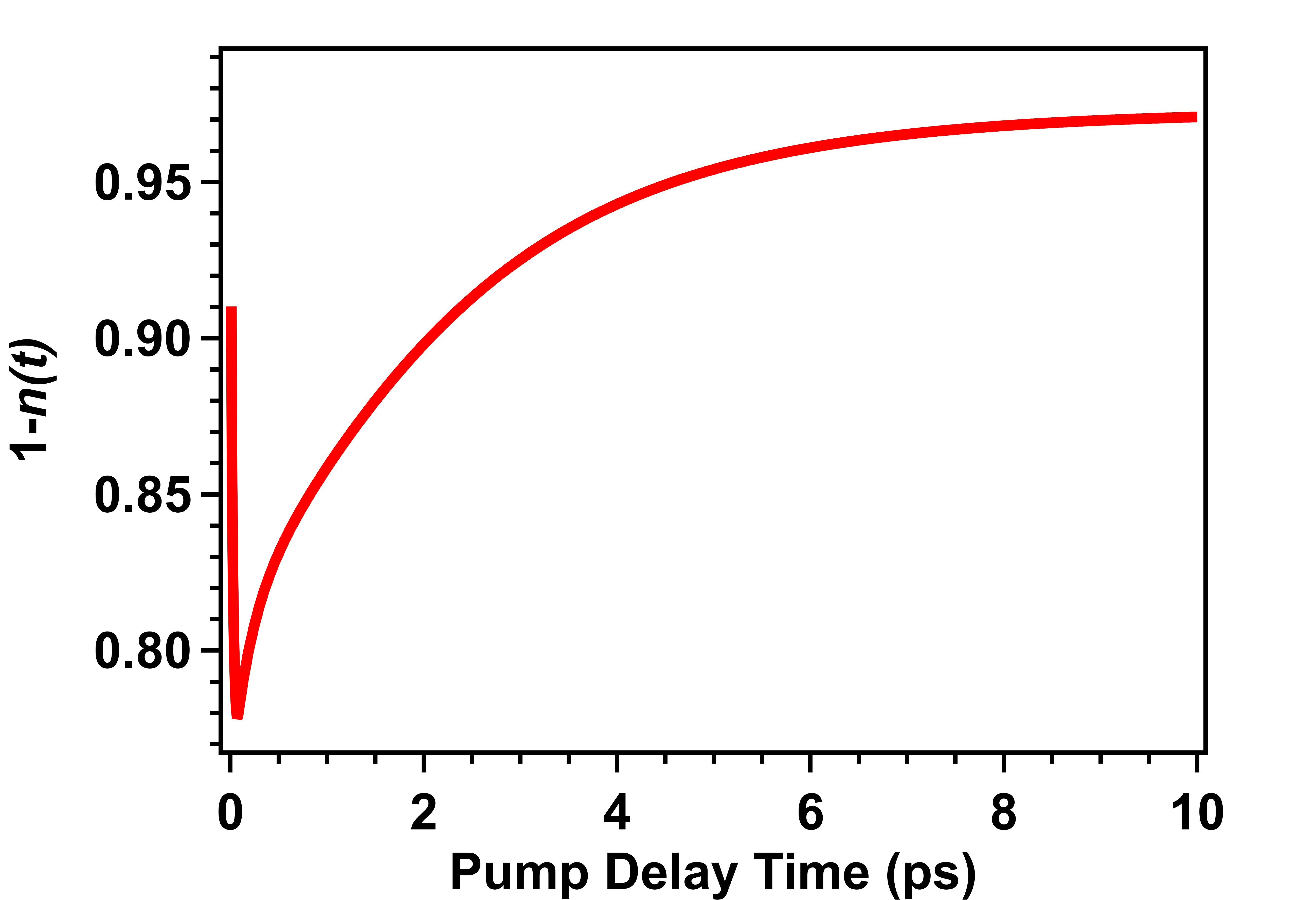}
    \caption{Calculated quantity (1 – population) of the \ce{MoS2} conduction band minimum as obtained from solving the kinetic equation \ref{differential} in section \ref{model}.}
    \label{fig:S4}
\end{figure}

\section{\label{details}Sample preparation, laser setup, and azimuthal-dependent FS-SFG pump-probe measurement}
\subsection{Sample preparation}
The sample preparation is based on the commonly used micromechanical exfoliation and has been reported elsewhere\cite{pollmann_large-area_2021}. The \ce{SiO2}/Si substrate was first cleaned and stored in ethanol before use. A Ti adhesive layer (5 nm) was grown by e-beam sputtering in a PVD system followed by deposition of a Au layer (25 nm) on the clean \ce{SiO2}/Si substrate. The base pressure of the PVD-chamber was set to $1.5\times10^{-5}$ mbar during deposition. After removing the freshly produced Au surface from the vacuum chamber, we immediately performed the conventional micromechanical exfoliation of \ce{MoS2}, resulting in a large area covered with a \ce{MoS2} monolayer. 

\subsection{Laser setup and azimuthal-dependent FS-SFG measurement}
We utilized a laser system consisting of a Ti: Sapphire oscillator (Vitara, Coherent) and a regenerative amplifier (Legend Elite Due HE + USP, Coherent). 
A portion of the regenerative amplifier output drove a commercial optical parametric amplifier (TOPAS-Prime, Light Conversion), where the signal and idler output were combined using a collinear difference frequency generation scheme, resulting in the production of infrared (IR). 
The narrow-band visible pulse was generated using an air-spaced etalon from SLS Optics Ltd. 
Both the visible beam (0.8 $\mu$J/pulse, full width at half maximum, FWHM: 0.003 eV, 1 kHz) and the tunable infrared beam ( below 1 $\mu$J/pulse, FWHM: 0.07 eV, 1 kHz) were propagated in the \textit{X-Z} plane and had incident angles of 64$^{\circ}$ and 46$^{\circ}$, respectively.
The pump beam (0.8 $\mu$J/pulse, FWHM: 0.051 eV, 1 kHz) was in the same incidence plane and had a incident angle of 59$^{\circ}$.
The sample was positioned randomly on a rotational stage, and while measuring the azimuthal dependence, the beam configuration remained the same while the sample was rotated relative to the surface normal.
The SFG signal was dispersed in a HRS-300 spectrograph from Princeton Instruments and projected onto a PI-MAX4 emICCD camera, also manufactured by Princeton Instruments.
The integrated SFG intensity was plotted as a function of azimuthal angle to guide the acquisition of FS-SFG spectrum at specific azimuthal angles, e.g., the angle with maximum intensity.
After the azimuthal angle was fixed at where the SFG signal has maximum intensity, the time-dependent FS-SFG spectra were collected with different pump delay times.
To capture a broad FS-SFG spectrum covering the band edge at \textbf{\textit{K}} point, the center frequency of the infrared beam was set from 3100 to 4300 nm.
The background was collected using the same settings by blocking the infrared beam.
Finally, the background-free FS-SFG spectrum was integrated and then normalized to the averaged FS-SFG spectra at equilibrium.   

\section{\label{resolution}Determine the $\sigma$ of the Gaussian function for fitting}
The temporal resolution of the experiment is determined by the pump-probe cross-correlation. 
As the probe beam utilized in this experiment is SFG, and currently, there is no feasible way to directly mearsure the pump-SFG cross-correlation, the pump-IR cross-correlation was instead obtained as a reference to estimate the temporal resolution of the pump-probe experiment.
As illustrated in Fig. \ref{fig:S2} of the Supplemental Material, the pump-IR cross-correlation trace measured on the Au mirror reveals a FWHM of 127 fs.
Accordingly, the $\sigma$ of this cross-correlation trace is 54 fs.
The pump beam was centered at 1.56 eV with a FWHM of 0.051 eV. 
The minimum pulse duration for an ideal transform-limited pump pulse is 36 fs.
Given that the probe beam, i.e. SFG pulse, has a shorter pulse duration than the IR pulse due to the time-energy uncertainty principle.
Therefore, the FWHM of the pump-SFG cross-correlation should exhibit a value less than 127 fs, which represents the upper limit of the $\sigma$ at 54 fs.
Consequently, we selected 50 fs as the $\sigma$ of the Gaussian function for fitting the dynamics of both \ce{MoS2} on Au and pure Au.

\section{\label{model}A simple kinetic model}
The transfer of charge between Au and \ce{MoS2} is analyzed by using a simple kinetic model. 
Previous first-principles calculations by us (Ref. \cite{yang_isolating_2024}, supplementary Fig. S5) demonstrated that the conduction band minimum (CBM) of  \ce{MoS2} monolayer when adsorbed on Au in a ($\sqrt{3}\times\sqrt{3}$) geometry forms an electronic resonance above the Fermi energy $E_F$ of Au(111). 
The CBM is located at $E_{CBM}$  = 0.09 eV with respect to $E_F$ in density functional calculations including spin-orbit coupling.  
While the wavefunctions of Au and \ce{MoS2} in this energy range are delocalized over both materials, in the kinetic model we single out the non-equilibrium population of the \ce{MoS2} orbitals near the CBM by defining a population $n(t)$, with $0 < n(t) < 1$, very much in the spirit of a Newns-Anderson model \cite{newns_self-consistent_1969}. 
To account for Coulomb repulsion between the electrons populating the CBM, we introduce an effective on-site repulsion $U$ in \ce{MoS2}. 
This allows to account for the experimentally observed non-equilibrium shift of the spectral centroid of about 1 meV while charging the \ce{MoS2} layer, see Fig. \ref{fig:S3}. 
In mean-field theory, we simply have $E_{CBM}(t)$ = $E_{CBM}$ + $U n(t)$.

In our model, the electrons excited in Au can tunnel into and out-off the CBM of \ce{MoS2} with a rate $R = 5\times10^{12}$ $s^{-1}$, i.e. the typical time for tunneling is 200 fs. 
This is on the same order of magnitude as the $\sim50$ fs reported previously from tr-ARPES experiments \cite{cabo_observation_2015}. 
For electrons excited high above the Fermi energy, as they occur shortly after the excitation, it is easier to overcome the energy barrier between Au and \ce{MoS2} monolayer. Therefore we use an enhanced rate for populating the \ce{MoS2} CBM at the sub-picosecond scale, $R_{in}(t) = R\cdot$max($1, 5-t/0.02$ps).

Moreover, the rate for tunneling in (out) depends on the number of occupied (unoccupied) states in the Au film which are described by a Fermi distribution function with variable temperature $T(t)$. 
Thus, the population $n(t)$ follows a first-order differential equation that can be written as
\begin{equation}
    \frac{dn(t)}{dt} = -R\cdot n(t) \Bigl( 1 - f(E_{CBM}(t),T(t)) \Bigr) + R_{in}(t) \Bigl( 1 - n(t) \Bigr) f(E_{CBM}(t),T(t)).
    \label{differential}
\end{equation}
Here, the Fermi distribution function $f(E_{CBM}(t),T(t))$ describes the non-equilibrium occupation of the electronic states in Au subsequent to the excitation by the pump laser pulse. 
We describe the time dependence of the electronic temperature entering into $f$ by a bi-exponential decay, 
\begin{equation}
    T(t) = T_0 + T_1e^{\frac{-t}{\tau_1}} + T_2e^{\frac{-t}{\tau_2}}.
\end{equation}

Using data for the fluence in the pump laser pulse and the electronic specific heat of the Au film, we arrive at the conclusion that the electronic system is heated to an initial temperature $T_0$ + $T_1$ + $T_2$ = 4000 K. 
To fix the other parameters in the expression for $T(t)$, we make the assumption that the break-in of the SFG at the pumped pure Au film directly reflects the (nearly homogeneous) electronic temperature. 
This gives us $\tau_1$ = 0.01 ps, $\tau_2$ = 2.31 ps ,  $T_1$ = 3259 K ,  $T_2$ = 446 K, while $T_0$ = 295 K is the room temperature at which the experiments are carried out.

We solve the kinetic equation with the initial condition $n(t=0)$ = 0. The quantity 1 – $n(t)$ is displayed in Fig. \ref{fig:S4}. 
The maximum population ($n_{max}$) reached is $\sim 0.22$, which corresponds to a drop of the SFG signal of \ce{MoS2} to about 0.78 = 1 – $n_{max}$, in good agreement with the experimental observation (0.75). 
With the knowledge of the spectral centroid shift of about 1 meV, we can determine the value of $U$ = 0.033 eV. 
The decay of the population $n(t)$, and thus the recovery of the SFG signal, is almost independent of the tunneling rate $R$, but rather follows the time constant $\tau_2$ of the electronic population of Au. 
An double-exponential fit to the curve shown in Fig. \ref{fig:S4} yields a second decay time constant of 2.31 ps , which is the same as the $\tau_2$ = 2.31 ps for the electrons in Au. 
The spectral centroid shift, which is in our model simply given by $U n(t)$, follows the same temporal behavior as $n(t)$ itself. 
Thus, the model consistently explains the experimental findings. 


%